\documentclass[12pt,preprint]{aastex}
\def\be{\begin{equation}}
\def\ee{\end{equation}}

\def\ergs{{\rm\,erg\,s^{-1}}}
\def\msun{M_{\odot}}

\def\ergs{\rm \,erg\,s^{-1}}
\def\be{\begin{equation}}
\def\ee{\end{equation}}
\catcode`\@=11 
\def\@versim#1#2{\vcenter{\offinterlineskip
        \ialign{$\m@th#1\hfil##\hfil$\crcr#2\crcr\sim\crcr } }}
\def\lsim{\mathrel{\mathpalette\@versim<}}
\def\gsim{\mathrel{\mathpalette\@versim>}}
\def\mpy{M_\odot \ {\rm yr^{-1}}}

\shorttitle{Nonthermal electrons in RIAF Models of Sgr A*}
\shortauthors{Yuan, Quataert, \& Narayan}

\begin{document}

\title{Nonthermal Electrons in Radiatively Inefficient Accretion Flow Models of Sagittarius A*}

\author{Feng Yuan}
\affil{Harvard-Smithsonian Center for Astrophysics, 60 Garden Street,
Cambridge, MA 02138}
\email{fyuan@cfa.harvard.edu}

\author{Eliot Quataert}
\affil{Astronomy Department, 601 Campbell Hall, University of California,
      Berkeley, CA 94720}
\email{eliot@astron.berkeley.edu}

\author{Ramesh Narayan}
\affil{Harvard-Smithsonian Center for Astrophysics, 60 Garden Street,
Cambridge, MA 02138}
\email{narayan@cfa.harvard.edu}

\begin{abstract}
We investigate radiatively inefficient accretion flow models for Sgr
A*, the supermassive black hole in our Galactic Center, in light of
new observational constraints.  Confirmation of linear polarization in
the submm emission argues for accretion rates much less than the
canonical Bondi rate. We consider models with low accretion rates and
calculate the spectra produced by a hybrid electron population,
consisting of both thermal and nonthermal particles. The thermal
electrons produce the submm emission and can account for its linear
polarization properties.  As noted in previous work, the observed
low-frequency radio spectrum can be explained if a small fraction
($\approx 1.5$\%) of the electron thermal energy resides in a soft
power-law tail. In the innermost region of the accretion flow,
turbulence and/or magnetic reconnection events may occasionally
accelerate a fraction of the electrons into a harder power-law tail.
We show that the synchrotron emission from these electrons, or the
Compton up-scattering of synchrotron photons by the same electrons,
may account for the X-ray flares observed by {\it Chandra}. 

\end{abstract}


\section{Introduction}

There is compelling evidence that the center of our Galaxy hosts a
massive black hole (BH) with mass $M=2.6 \times 10^6 \msun$ (e.g.,
Sch\"odel et al. 2002; Ghez et al. 2003).  The inferred location of
the black hole is coincident with the energetic radio source Sgr A$^*$
(e.g., Melia \& Falcke 2001),  which has been intensively studied
since its discovery. The radio spectrum of Sgr A* consists of two
components which dominate below and above 86 GHz, respectively.  The
component below 86 GHz has a spectrum $F_{\nu} \propto \nu^{0.2}$,
while the high frequency component, the ``submm bump'', has a spectrum
$F_{\nu} \propto \nu^{0.8}$ up to $\sim 10^3$ GHz (Zylka, Mezger \&
Lesch 1992; Serabyn et al. 1997; Falcke et al. 1998; Zhao et
al. 2003).  At yet higher frequencies, in the IR, there are strong
upper limits to the flux, indicating that the spectrum cuts-off steeply (e.g.,
Hornstein et al. 2002).  The IR limit plus an X-ray detection from
{\it Chandra} (Baganoff et al. 2001, 2003) indicate that Sgr A$^*$ is
quite dim overall, with a bolometric luminosity of only $L \approx
10^{36}$ ergs s$^{-1}$ $\approx 3 \times 10^{-9} L_{\rm Edd}$.  Most
of this luminosity is radiated in the submm bump.

The proximity of the Galactic Center allows one to determine
observationally the dynamics of gas quite close to the BH, providing
unique constraints on theoretical models of the accretion flow.  A
canonical estimate for the rate at which a BH gravitationally captures
surrounding gas is given by the spherical 
accretion model of Bondi (1952).  This
model can be applied directly to Sgr A$^*$ (see Melia 1992 for an early
application and Melia, Liu, \& Coker 2001 for a current version
of the spherical accretion model).  {\it Chandra} observations of the
Galactic Center detect extended diffuse emission within $1-10''$ of
the BH (Baganoff et al. 2003).  This emission likely arises from hot
gas produced when the winds from massive stars collide and shock.
Interpreted as such, the inferred gas density and temperature are
$\approx 130$ cm$^{-3}$ and $\approx 2$ keV about $1''$ from the BH.
A remarkable coincidence is that, given the measured BH mass and
measured ambient temperature, the sphere of influence of the BH
(defined by the Bondi accretion radius) is $R_{acc} \approx
GM/c_\infty^2 \approx 0.04 \ {\rm pc} \approx 1''$, comparable to {\it
Chandra's} resolution!  The Bondi accretion rate is thus very well
determined: $\dot M_B \approx 10^{-5} M_\odot$ yr$^{-1}$.  If gas were
accreting at this rate onto the BH via a thin accretion disk (Shakura
\& Sunyaev 1973), a model that has been extensively and successfully
applied to {luminous} accreting sources (e.g., Koratkar \& Blaes
1999), the expected luminosity would be $L \approx 0.1 \dot M_B c^2
\approx 10^{41}$ ergs s$^{-1}$.  This is larger than the observed
bolometric luminosity of Sgr A* by a factor of $\sim 10^5$.  The
observations thus favor accretion models in which very little of the
gravitational potential energy of the inflowing gas is radiated, i.e.,
$L \ll \dot M_{B} c^2$.  We will refer to such models as radiatively
inefficient accretion flows (RIAFs).  The distinction between a Bondi
accretion flow and a RIAF is that the former describes a spherical
flow with no angular momentum while the latter describes a hot
quasi-spherical rotating accretion flow with viscosity.

Advection-dominated accretion flows (ADAFs) are an
analytically-motivated model for the dynamics of RIAFs (e.g., Narayan
\& Yi 1994).  In such models, $\dot M \approx \alpha \dot M_B$ (where
$\alpha\sim0.1$ is the dimensionless viscosity parameter) and $\rho
\propto r^{-3/2}$, as in spherical Bondi accretion.  A number of
studies have shown that ADAF models accreting at the observationally
inferred rate could roughly account for the spectrum and luminosity of
Sgr A* (Narayan et al. 1995, 1998; Manmoto et al. 1997; Mahadevan
1998; \"Ozel, Psaltis \& Narayan 2000; Narayan 2002). 
 The key requirement in these models is that the
fraction of the turbulent energy heating the electrons, $\equiv
\delta$, has to be quite small, $\delta \lsim 0.01$, in order to
explain the low luminosity of Sgr A*; the corresponding electron
temperature close to the BH is $T_e \sim 10^{10}$ K $\ll T_p \sim
10^{12}$ K.

In this paper we reexamine RIAF models for the spectrum of Sgr A*.  We
are motivated both by important new observational results on the
emission from Sgr A* and by theoretical work which shows that the
analytical ADAF model does not accurately describe the dynamics of
RIAFs.  We first summarize the new observational constraints.

Recent radio observations show that Sgr A* has no linear polarization
between 1.4 and 112 GHz with limits of 0.1 to 2\% (Bower et
al. 1999a,b, 2001), while the source has measurable circular
polarization between 1.4 and 43 GHz (Bower, Falcke, \& Backer
1999). At higher frequencies, JCMT observations at 150, 220, 375 and
400 GHz indicate that Sgr A* is linearly polarized at a level of $\sim
10\%$ (Aitken et al. 2000).  This result was confirmed by the BIMA
array, which has a much better angular resolution --- Bower et
al. (2003) detected a linear polarization of $7.2 \pm 0.6 \%$ at 230
GHz.  This observation alone places an upper limit on the rotation
measure of $2\times10^6 \ {\rm rad~m}^{-2}$.  In an ADAF model, as
originally formulated, the region where the (linearly polarized)
submm-bump is predicted to originate has such a large electron density
that the rotation measure would be many orders of magnitude larger
than the above observational upper limit (Quataert \& Gruzinov 2000;
Agol 2000).  Assuming that the magnetic field does not undergo many
reversals along the line of sight (see Ruszkowski \& Begelman 2002),
the polarization data thus impose a serious constraint on models of
the accretion flow.

At X-ray wavelengths, {\em Chandra} has convincingly detected an X-ray
source coincident with Sgr A* (to within $\approx 0.2'',$ which is
$\approx 0.01$ pc or $\approx 3 \times 10^4$ $R_S$ at the Galactic
Center, where $R_S$ is the Schwarzschild radius of the BH).  {\it
Chandra} observations show that this X-ray source comes in two states.
In the {\em quiescent state}, the absorption-corrected 2-10 keV
luminosity is 2.2$^{+0.4}_{-0.3} \times 10^{33} {\rm erg ~s}^{-1}$.
The spectrum is soft and is well fitted by an absorbed power-law model
with photon index $\Gamma=2.2^{+0.5}_{-0.7}$.  A large fraction of the
X-ray flux comes from an extended region with diameter $\approx
1.4^{\arcsec}$. Comparisons between two observations separated by
about a year show that this component remains constant (Baganoff et
al. 2003). In the {\em flare state}, the luminosity of Sgr A*
increases by a factor of few to 45 over a timescale of minutes to
hours; the short timescale argues that the emission arises quite close
to the BH, roughly within $\lsim 10-100 R_S$.  For the strongest
flare, the luminosity is $\approx 10^{35}\ergs$, the timescale is
about 3 hours, and the spectrum is hard, $\Gamma=1.3^{+0.5}_{-0.6}$
(Baganoff et al. 2001). Further observations by both {\em XMM-Newton}
(Goldwurm et al. 2003) and {\em Chandra} (Baganoff 2003a) indicate
that rapid and intense X-ray flares, with a factor of five or more
increase in luminosity, are relatively common and occur roughly once
per day. Such flares provide important constraints on the gas dynamics
close to the BH.

To search for counterparts to the X-ray flares, a multiwavelength
campaign was conducted, including radio, submm, IR, and X-ray
observations (Baganoff 2003a). The X-ray flares
are apparently not accompanied by large variations at longer wavelengths,
though there may be some evidence for variations at mm wavelengths 
at the level of a few tens of percent.

In addition to these new observational constraints, there has been a
significant change in the theoretical understanding of RIAFs over the
past few years.  Most importantly, global, time-dependent, numerical
simulations reveal that the structure of the flow is very different from
the original self-similar ADAF prediction of Narayan \& Yi (1994).  The key
result from nearly all such simulations is that {$\dot M \ll \dot
M_{B}$, i.e., very little mass available at large radii actually
accretes onto the black hole}; most of it is lost to a magnetically
driven outflow or circulates in convective motions (e.g., Stone,
Pringle, \& Begelman 1999; Igumenshchev \& Abramowicz 1999;
Igumenshchev et al. 2000; Stone \& Pringle 2001; Hawley \& Balbus
2002; Igumenshchev et al. 2003). Another way to state this result is
that for a given gas density at a large distance from the black hole
(e.g., measured by {\it Chandra} on 1'' scales), the density close to
the BH is much {less} than the ADAF or Bondi predictions.  This
theoretical result is consistent with the linear polarization
detection from Sgr A*, which argues strongly for low gas densities
close to the BH and thus low $\dot M \ll \dot M_B$.

Quataert \& Narayan (1999) showed that RIAF models with $\dot M \ll
\dot M_B$ could account for the basic observed properties of Sgr A*.
To produce the correct amount of emission, however, their models required that
the electrons in the accretion flow should be hotter than in ADAF
models ($\sim 10^{11}$ K close to the BH rather than $10^{10}$ K).
Stated another way, $\delta$, the fraction of the turbulent energy
that heats the electrons, must be larger.  Given the theoretical
uncertainties in how electrons are heated in the accretion flow, this
is very plausible (e.g., Quataert \& Gruzinov 1999).
However, in the work of Quataert \& Narayan (1999), 
old X-ray spectral data were used and
the accretion rate at the outer boundary was treated as a free parameter.
Given the considerable progress in the field during the last few years,
it is neccesary to check whether a RIAF model can fit the new data,
specifically, (i) the {\it Chandra} X-ray spectrum, (ii) the measured 
gas density at the accretion radius (from {\it Chandra}), and (iii) the gas
density near the BH (from radio polarization data).
This is the goal of the present paper.

An additional issue is the electron distribution function.
Since the inflowing gas is collisionless, processes such as MHD
turbulence, reconnection, and weak shocks can accelerate electrons and
generate a nonthermal tail at high energies in the electron
distribution function.  The effect of such nonthermal electrons on the
synchrotron spectrum of Sgr A* was investigated by Mahadevan (1998)
and \"Ozel et al. (2000) (see Wardzi\'nski \& Zdziarski 2001 for a
discussion of the effect of a power-law tail on synchrotron and
inverse Compton emissions in luminous systems).  They found that the
low-frequency radio spectrum, which was under-predicted in earlier
ADAF models (e.g., Narayan et al. 1998), could be explained quite
well, if roughly $1\%$ of the steady state electron energy is in
nonthermal electrons\footnote{Liu \& Melia (2001) reached a similar 
conclusion: they showed that the combination of centimeter and X-ray data
preclude the possibility of producing the observed 1.36 GHz radio flux via
thermal synchrotron emission in a bounded accretion flow and that nonthermal
electrons are needed}. However, the calculations were based on a model
with a high accretion rate in the inner region of the accretion flow,
$\dot{M}\approx 3 \times 10^{-6} \msun$ yr$^{-1}$, which is about two
orders of magnitude larger than the upper limit on the accretion rate
implied by the radio linear polarization measurements. Therefore, a
secondary goal of the present paper is to examine how the predictions
for the radio emission are modified when the density in the inner
region of the accretion flow is much lower than previously
considered.



In the calculations reported here we focus in particular on non-thermal
electrons, which can account for the low frequency radio emission and
which may also be responsible for the X-ray flares seen by {\it
Chandra}.  In \S 2, we present our method for treating a nonthermal
electron distribution function in the accretion flow. In \S3, we
present our results of fitting the quiescent spectrum of Sgr A*.  Then, in \S4,
we consider possible origins of the X-ray flares observed in Sgr A*.
We conclude with a summary and discussion in \S5.

\section{Methodology}

\subsection{Dynamics of the Accretion Flow}

To take into account the role of outflows/convection in modifying the
density profile of the RIAF, we write the dependence of $\dot{M}$ on
radius as follows (Blandford \& Begelman 1999),

\be \dot{M}=-4\pi R H \rho v = \dot{M}_{\rm out}\left(\frac{R}{R_{\rm
out}} \right)^s, \ee where $\dot{M}_{\rm out}$ is the mass accretion
rate at the outer boundary of the flow ($\sim \alpha \dot M_B$),
$R_{\rm out} \approx R_{acc} \approx 10^5 R_s$ is the ``outer radius''
of the flow (the Bondi accretion radius), and $H, \rho,$ and $v$ are
the scale height, mass density, and radial velocity, respectively;
we set $H=c_s/\Omega_{\rm K}$ where $c_s$ is the sound speed and $\Omega_{\rm K}$
is the Keperian angular velocity.
The parameter $s$ describes how the density profile and accretion rate
are modified.  The net accretion rate onto the BH is $\dot M_{in}
\approx \dot M_{out} (R_{in}/R_{out})^s$; in a self-similar flow this
would correspond to a density profile $\rho(r) \propto r^{-3/2 + s}$.

We assume that a large fraction $\delta$ of the turbulent energy
directly heats the electrons. Assuming for now that the electrons are
thermal, the electron energy equation is given by \be \rho v
\left(\frac{d \varepsilon_e}{dr}- {p_e \over \rho^2} \frac{d
\rho}{dr}\right) =\delta q^++q_{ie}-q^-, \ee where $\varepsilon_e$
is the internal energy of electrons per unit mass of the gas, $p_e$ is
the pressure due to electrons, $q_{ie}$ is the
Coulomb energy exchange rate between electrons and ions, $q^-$ is the
electron cooling rate, and $q^+$ is the net turbulent heating rate.
The energy equation for the ions is then given by \be \rho v
\left(\frac{d \varepsilon_i}{dr}- {p_i \over \rho^2} \frac{d \rho}{dr}
\right) =(1-\delta)q^+-q_{ie}=-(1-\delta)\alpha p r
\frac{d\Omega}{dr}-q_{ie}. \ee The other two equations, namely the
radial and azimuthal components of the momentum equation, are \be v
\frac{dv}{dr}=-\Omega_{\rm k}^2 r+\Omega^2 r-\frac{1}{\rho}\frac
{dp}{dr}, \ee \be v(\Omega r^2-j)=\alpha r \frac{p}{\rho}. \ee Here,
$j$ is the specific angular momentum of the gas accreted into the BH,
$\Omega$ denotes the angular velocity, and $p=p_i+p_e$ is the total
gas pressure (electron plus ion). We
adopt the Paczy\'nski \& Wiita (1980) potential to mimic the geometry
of a Schwarzschild black hole.  

We solve the above set of equations,
with appropriate boundary conditions (e.g., Yuan 1999; Yuan et
al. 2000), to obtain the flow characteristics such as ion and electron
temperatures, density, etc.  We should note that the equations solved
here are not fully consistent since they do not account for any energy
or angular momentum that is transported by convection or is lost to an
outflow.  We expect that this will not significantly change our
results since the primary effect of outflows/convection is to modify
the density profile of the flow, which we have accounted for (small
additional uncertainties in, e.g., the temperature, are absorbed into
the large uncertainties in the heating parameter $\delta$).

\subsection{Electron Distribution Function}

We consider a hybrid distribution of electrons at each radius in the
accretion flow. The thermal distribution is \be n_{\rm
th}(\gamma)={N_{\rm th}\gamma^2\beta \exp (-\gamma/\theta_e) \over
\theta_e {\rm K}_2(1/\theta_e)}, \ee where $\gamma$ is the electron
Lorentz factor, and $\theta_e\equiv kT_e/m_e c^2$ is the dimensionless
electron temperature.  The power-law distribution is described by
\begin{mathletters}
\be
n_{\rm pl}(\gamma)=N_{\rm pl}(p-1)\gamma^{-p}, \hspace{1cm}
 \gamma_{\rm min} \le
\gamma \le \gamma_c,
\ee
\be
n_{\rm pl}(\gamma)=N_{\rm pl}(p-1)\gamma_c \gamma^{-p-1}, \hspace{1cm}
 \gamma_{\rm c} \le
\gamma \le \gamma_{\rm max}, \ee
\end{mathletters}
where $\gamma_{\rm min}, \gamma_{\rm max}$ and $\gamma_{\rm c}$ are
the minimum, maximum, and the ``cooling break'' Lorentz factors,
respectively. From the global solution for the accretion flow
structure, we obtain the strength of the magnetic field, $B$, at
each radius, assuming that the magnetic energy density is a fixed
fraction $\beta^{-1}$ of the thermal energy density (an equipartition
prescription).  The ``cooling break'' Lorentz factor $\gamma_c$ is
then determined by the condition \be t_{\rm cool} \equiv
\frac{3}{4}\frac{8\pi m_ec}{\sigma_{\rm T}\gamma_c
\beta_e^2B^2}=t_{\rm accretion}\equiv \frac{R}{|v|}. \ee Usually, the
cooled electrons have an energy index $(p+1)$ as written in equation
(7b) (Throughout the paper, $p$ denotes the spectral index of the
{\em injected} electrons in the power-law energy distribution). 
 However, if $p < 1$, the ``cooled'' electron distribution
function goes as $n(\gamma) \propto \gamma^{-2}$, not $n(\gamma)
\propto \gamma^{-p-1}$.

We calculate the values of $\gamma_{\rm min}$ and $N_{\rm pl}$ as
follows.  We assume that the {\it injected} energy in nonthermal electrons
is equal to a fraction $\eta$ of the energy in thermal electrons; as a
fiducial model, we take $\eta$ to be independent of radius (but see \S4
where we consider X-ray flares).  The energy density of thermal
electrons at temperature $\theta_e\equiv kT_e/m_ec^2$ is (Chandrasekhar
1939)\be u_{\rm th}=a(\theta_e)N_{\rm th}m_ec^2\theta_e, \ee where the
quantity \be
a(\theta_e)\equiv\frac{1}{\theta_e}\left[\frac{3K_3(1/\theta_e)+K_1(1/\theta_e)}
{4K_2(1/\theta_e)}-1\right] \ee varies from $3/2$ for nonrelativistic
electrons to $3$ for fully relativistic electrons, and $K_n$ are
modified Bessel functions of the $n$th order. The energy density of
power-law electrons is \be u_{\rm pl}\approx N_{\rm
pl}m_ec^2\frac{p-1}{p-2}\gamma_{\rm min}^{2-p} \ee for $p > 2$. So the
number density of power-law electrons is determined by $u_{\rm
pl}=\eta u_{\rm th}$, which gives \be N_{\rm
pl}=\frac{p-2}{p-1}\gamma_{\rm min}^{p-2}\eta a(\theta_e)
\theta_eN_{\rm th}. \ee We obtain a second relation from the condition
that the power-law distribution should match smoothly onto the thermal
distribution: \be n_{\rm th}(\gamma_{\rm min})=n_{\rm pl}(\gamma_{\rm
min}). \ee This condition is natural since the nonthermal electrons are
presumably accelerated out of the thermal pool. 
We solve equations (12) and (13) to obtain $N_{\rm pl}$ and
$\gamma_{\rm min}$ simultaneously as a function of radius.

When we consider models for flares (\S 4), we have $p < 2$. In this case, 
the formula corresponding to eqs. (11)-(12) are
\be
u_{\rm pl}\approx N_{\rm pl}m_ec^2\frac{p-1}{2-p}\gamma_{\rm max}^{2-p},
\ee
and 
\be
N_{\rm pl}=\frac{2-p}{p-1}\gamma_{\rm max}^{p-2}\eta a(\theta_e) \theta_eN_{\rm th}.
\ee
The value of $\gamma_{\rm max}$ depends on the details of
electron acceleration which are not understood. 
We treat it as a free parameter,
and adopt the minimum $\gamma_{\rm max}$ required to 
fit the X-ray spectrum of the flare
detected by {\em Chandra}.

\subsection{Synchrotron and Inverse-Compton Emissions}

After determining the distribution of both thermal and power-law
electrons, we calculate their radiation. Following Mahadevan, Narayan \& Yi
(1996), the optically thin synchrotron emissivity of a relativistic
Maxwellian distribution of electrons is \be j_{\nu,{\rm
th}}=\frac{4\pi N_{\rm th}e^2}{\sqrt{3}cK_2(1/\theta_e)} \nu M(x_M),
\ee where \be x_M\equiv \frac{2\nu}{3\nu_b\theta_e^2}, \hspace{1cm}
\nu_b\equiv \frac{eB}{2\pi m_e c},\ee with $M(x_M)
$ given by \be
M(x_M)=\frac{4.0505}{x_M^{1/6}}\left(1+\frac{0.40}{x_M^{1/4}}+\frac{0.5316}
{x_M^{1/2}}\right) \exp (-1.8899x_M^{1/3}).  \ee The synchrotron
absorption coefficient $\alpha_{\rm th}$ is related to the emissivity
via Kirchoff's law.

For the emissivity of power-law electrons we use the following
expressions (Westfold 1959; Blumenthal \& Gould 1970), which are more
exact than the analytically integrated formula given in \"Ozel et
al. (2000), especially for frequencies near the
``cutoff'' and the ``break''
in the power-law distribution:
\be j_{\rm pl}=\frac{1}{2}(p-1)N_{\rm
pl}\left(\frac{\sqrt{3}e^2}{2c}\right)\nu_b
\left(\frac{2\nu}{3\nu_b}\right)^{(1-p)/2}H(p)\left[G(x_2)-G(x_1)\right],
\ee where $x\equiv \nu/\nu_c$, 
$\nu_c=(3/2)\nu_b\gamma^2$.
The quantities $x_1$ and $x_2$ correspond to the minimum and maximum
Lorentz factors of the power-law electrons. In the above equation,
\begin{eqnarray}
H(p)&=&\int({\rm sin}~\alpha)^{(1+p)/2}g(\alpha)d\Omega_{\alpha} \nonumber \\ &=&2\pi
\sqrt{\pi}\Gamma\left(\frac{p+5}{4}\right)/\Gamma\left(\frac{p+7}{4}\right),
\end{eqnarray}
\be
G(x_1)=\int^{\infty}_{x_1}x^{(p-1)/2}\int^{\infty}_{x}K_{5/3}(y)dydx.
\ee For the absorption coefficient of power-law electrons, we use
(\"Ozel et al. 2000) \be \alpha_{\nu,{\rm pl}}=C^{\alpha}_{\rm
pl}\eta\frac{e^2N_{\rm th}}{c}
a(\theta_e)\theta_e\left(\frac{\nu_b}{\nu}\right)^{(p+3)/2}\nu^{-1},
\ee with \be C^{\alpha}_{\rm
pl}=\frac{\sqrt{3\pi}3^{p/2}}{8}\frac{p-1}{m_e}\\
\frac{\Gamma[(3p+2)/12]\Gamma[(3p+22)/12]\Gamma[(6+p)/4]}
{\Gamma[(8+p)/4]}. \ee

We also consider the Comptonization of seed synchrotron photons, both
those produced by thermal electrons and those produced by power-law
electrons. The scattering electrons can also be either thermal or
power-law.  We adopt the method of Coppi \& Blandford (1990) to
calculate the Comptonization. This method uses a local approximation,
but it gives reasonable results when compared with the more careful
Comptonization calculations of Narayan, Barret, \& McClintock (1997).

\subsection{Radiative Transfer Calculation}

To calculate the spectrum, we adopt the ``plane parallel rays'' method
(Mihalas 1978; \"Ozel et al. 2000).  In this method, the equation of radiative
transfer for a time-independent, spherically symmetric flow is written
for plane parallel rays of varying impact parameters through the flow
and solved along these rays (Mihalas 1978).
We integrate the equation using the formal solution
and the approximate boundary conditions for a nonilluminated
atmosphere (Mihalas 1978). We carry out the integral along rays with
impact parameters up to $\sim 5000 R_s$, beyond which the electron
temperature is too low for significant synchrotron emission.  The
total flux is then obtained by integrating over all impact parameters.
See \"Ozel et al. (2000) for details.

\section{Quiescent State Spectra}

Figure 1 shows the spectral data on Sgr A* that we wish to explain.
The main challenges for a model are (i) to explain the submm bump and
the excess radio emission at low frequencies, (ii) to satisfy the
stringent upper limit in the infrared, and (iii) to explain the low
X-ray flux in quiescence (the lower ``bowtie'' in Fig. 1), the fact
that most of the quiescent X-ray flux is spatially resolved, and that it has a
relatively soft X-ray spectrum.  We wish to explain all these facts
while also satisfying the density measured by {\it Chandra} at the 
accretion radius, and the upper limit on the density of the gas near
the black hole as deduced from the linear polarization data of Aitken
et al. (2000) and Bower et al. (2003).

The solid line in Fig. 1 shows the spectrum corresponding to a RIAF
model of Sgr A*, for which the flow parameters vary with radius as
shown in Fig. 2.  The parameters in the model are $\alpha = 0.1$,
plasma $\beta = 10$, i.e., the magnetic energy density is $10\%$ of
the thermal energy density, $\dot{M}_{out} \approx 10^{-6} \mpy$, $s =
0.27$, $\delta = 0.55$, $p=3.5$, and $\eta=1.5\%$.

Fig. 1 shows that the model fits the spectrum of Sgr A* fairly well,
including the low-frequency radio data, the submm-bump, and the {\it
Chandra} X-ray emission.  The low frequency radio emission is
self-absorbed synchrotron emission from power-law electrons in the
accretion flow. As discussed by \"Ozel et al. (2000), this part of the
spectrum is determined by a very unusual source function, namely $S=
j_{\rm pl}/\alpha_{\rm th}$ (since at these frequencies $\alpha_{\rm
th} \gg \alpha_{\rm pl}$ while $j_{\rm pl} \gg j_{\rm th}$).
Mahadevan (1998) and \"Ozel et al. (2000) used a canonical ADAF model
to show that a small fraction of power-law electrons can explain the
low-frequency radio data in Sgr A*. We find that the same is true even
when $\dot{M}$ near the black hole is much less than in their
models. The IR emission in our model is due to optically-thin
synchrotron emission (and some SSC) by the same power-law electrons
that produce the radio emission.  By contrast, the submm-bump is
primarily produced by thermal electrons in the inner parts of the
RIAF.  Finally, the X-ray emission is thermal bremsstrahlung emission.
This emission originates at large radii far from the BH, and is
consistent with {\it Chandra} observations which indicate that the
source is extended, with a size of $1''$.  Moreover, because the Bondi
accretion radius, $R_{acc}$, is also $\approx 1''$, the observed
emission comes from the ``transition region'' between the ambient
medium and the accretion flow, where the gas is being gravitationally
captured by the central BH.  To account for this, we have used
Quataert's (2002) calculation of the bremsstrahlung spectra produced
by correctly matching the accretion flow onto the ambient medium at
radii $\approx R_{acc}$.

It is important to check whether the model in Fig. 1 can satisfy the
Faraday rotation measure constraint.  We find that ${\rm RM} \approx
10^7 \ {\rm rad \ m^{-2}}$ if we integrate through the equatorial
plane of the accretion flow, while ${\rm RM} \la 5 \times 10^5 \ {\rm
rad \ m^{-2}}$ at the region where most of the emission at 230 GHz
comes from if we integrate along the rotation axis. This large
difference arises because in the latter case we only see the hot
relativistic inner regions of the accretion flow (the ${\rm RM}$
produced by gas on the scales resolved by {\it Chandra} is only ${\rm
RM} \approx 3 \times 10^3 \ {\rm rad \ m^{-2}}$ assuming equipartition
mG magnetic fields).  It should also be noted that these estimates are
probably upper limits because they assume that the magnetic field is
fully coherent and points along the line of sight (while the magnetic
field in the accretion flow is actually predominantly toroidal).
Given the above uncertainties, we are reasonably consistent with Bower
et al.'s (2003) limit of ${\rm RM} \lsim 2 \times 10^6 \ {\rm rad \
m^{-2}}$.

As noted above, the submm emission in our models is due to thermal
electrons. Since the linear polarization of optically thick thermal
synchrotron emission from a uniform medium is suppressed by
$\exp(-\tau)$, where $\tau$ is the synchrotron optical depth, we also
need to check whether our model can roughly account for the magnitude
of the observed linear polarization ($\approx 10\%$).  In addition, an
interesting possibility is that this strong $\tau$
dependence could explain the observed rapid variation of linear
polarization with frequency (from $\approx 7\%$ at 230 GHz to $\lsim 2
\%$ at 112 GHz).  We have calculated the linear polarization produced
by the thermal electrons in our models; we include optical depth
effects and Faraday rotation using the formula in Pacholczyk (1970),
but have not calculated the conversion of linear to circular
polarization.  We assume that our line of sight is well out of the
equatorial plane of the accretion flow so that the path length through
the accretion flow at any radius $R$ is $\approx H(R)$, the scale
height.  Note that the Faraday rotation measure is a function of
radius (and is $\sim 10^6 \ {\rm rad \ m^{-2}}$ for the radii that
dominate the submm emission considered here).

In Fig. 3a we show $\tau$ as a function of radius for three
frequencies and in Fig. 3b we show by open circles the degree of
linear polarization as a function of frequency when Faraday rotation
is neglected.  At $\approx$ 400 GHz, the emission from all radii is
optically thin, and the emission is calculated to be $\approx 70 \%$
linearly polarized for a uniform B-field.  At $\approx 230$ GHz, the
maximal optical depth is $\tau \approx 2$, which suggests a factor of
10 suppression in the linear polarization.  In fact, however, emission
from radii with $\tau \lsim 1$ contributes a large fraction of the
flux, so we find that the net polarization (integrated over all radii)
should still be very high, $\approx 59 \%$.  At $\approx 112$ GHz, a
similar estimate yields a maximal polarization of $\approx 45 \%$.  We
thus conclude that thermal electrons can readily account for the
observed level of linear polarization from Sgr A*. This is
consistent with the work of Melia, Liu, \& Coker (2000) who use a
slightly different method to conclude that thermal electrons in their
spherical accretion models can account for the obseved level of linear
polarization. However, because of the contribution from $\tau \lsim 1$
regions, optical depth effects are unlikely to explain the large
change in polarization with frequency.  This must be due to other
effects. One possibility is Faraday depolarization by the accretion
flow itself (Bower et al 1999a, 2003; Quataert \& Gruzinov 2000). The
filled circles in Fig. 3b show the degree of polarization when Faraday
rotation is included.  The strong suppression of polarization at 112
GHz compared to 230 GHz (consistent with that observed) is because the
Faraday rotation angle is $\propto \nu^{-2}$ for the same rotation
measure, and because the emission at low frequencies comes from radii
where the rotation measure is actually somewhat larger. Other reasons
for the change in polarization could include a change in the magnetic
field geometry, or a change in the emission component (which in our
models only occurs at $\lsim 50$ GHz).

The model has several parameters, $\alpha$, $\beta$, $\dot M_{\rm
out}$, $s$, $\delta$, $p$, $\eta$, and it would be useful to
understand how the different parameters are determined and how well
they are constrained.  The viscosity parameter $\alpha$ and the
magnetic field parameter $\beta$ are known approximately from
numerical MHD simulations (e.g., Balbus \& Hawley 1998).
We have assigned reasonable values to these parameters,
$\alpha=0.1$, $\beta=10$, and have not considered variations.
Although the outer mass accretion rate $\dot M_{\rm out}$ is
technically an independent parameter, in practice it is degenerate
with $\alpha$ since the density of gas in the accretion flow scales
roughly as $\dot M/\alpha$.  Our choice of $\dot M_{out}/\alpha =
10^{-5} \mpy$ corresponds to a gas density of $\approx 130$ cm$^{-3}$
at $10^5 R_S$, in agreement with {\it Chandra} observations (see \S1).
Once $\alpha$ is fixed, there is little freedom in $\dot M_{\rm out}$.

The parameters $p$ and $\eta$ describe the distribution of the
power-law electrons.  The value of $\eta$ is determined by requiring
the predicted spectrum to agree with the radio data.  We have found
that a value $\sim 1.5\%$ gives a good fit to the data, more or less
independent of $p$, while values a factor of several larger or smaller
deviate noticeably from the data.  Figure 4 shows how the model shown
in Fig. 1 changes if we vary $p$, the slope of the power-law electron
distribution function, while keeping $\eta$ fixed at 1.5\%.  To
highlight the important changes, only the synchrotron emission from
thermal and power-law electrons is shown. Different values of $p$ give
almost the same fit to the low-frequency radio spectrum, as found by
\"Ozel et al. (2000). However, the predicted spectra in the IR and
X-ray bands show large variations.  From this comparison, we conclude
that $p \gsim 3.5$ is favored because otherwise the optically thin
power law tail becomes too prominent and violates the IR and/or X-ray
limits.

The parameters $s$ and $\delta$ behave as a pair.  For each value of
$s$, we determine $\delta$ by requiring the model to reproduce the
peak flux in the sub-mm bump.  For our baseline model with $s = 0.27$
and $\dot M_{\rm out} \approx 10^{-6} \mpy$, the accretion rate onto
the BH is $\dot M \approx 4 \times 10^{-8} \mpy$.  As we showed above,
this model is reasonably consistent with the RM constraint.  One way
to decrease RM even more is to decrease the accretion rate onto the BH
by increasing the power-law index $s$, defined in equation (1).
Fig. 5 shows the effect of varying $s$ (adjusting $\delta$
appropriately).  The dot-dashed line is the same as the dot-dashed
line in Fig. 1, i.e., it corresponds to $s=0.27, \delta=0.55$.  The
dashed line is for $s=0.1$ and $\delta=0.1$. This model fits the
spectrum very well, but its rotation measure is too large: ${\rm RM}
\sim 3 \times 10^8 \ {\rm rad \ m^{-2}}$, which implies that the
polarization would be completely suppressed at frequencies of order a
few hundred GHz.  In fact, this model is similar to the canonical ADAF
model (Narayan et al. 1998; \"Ozel et al. 2000), which is inconsistent
with the radio polarization data.  Lastly, the dotted line is for
$s=0.4$ and $\delta=1.0$.  This model significantly overpredicts the
flux in the submm bump over the frequency range $10^{10.5}-10^{11.5}$
Hz, as has been shown earlier by Quataert \& Narayan (1999) and Yuan,
Markoff, \& Falcke (2002).  In addition, because the RM is very low,
the polarization would be very high even at low frequencies, in
conflict with the observations.

\section{X-ray Flares}

The X-ray flares are the most dramatic result from the {\it Chandra}
observations of Sgr A*.  In this section we examine models for these
flares within the context of RIAFs.  Markoff et al. (2001) showed that
the flares are probably due to enhanced electron heating or
acceleration, rather than a change in the accretion rate onto the BH;
otherwise there is too much variation at lower frequencies compared to
the observations.  A possible analog of the {\it Chandra} flares are
solar flares, in which magnetic energy is converted into thermal
energy, accelerated particles, and bulk kinetic energy, giving rise to
a burst of radiation (e.g., Priest \& Forbes 2000).  Although it is empirically
established that particle heating and acceleration are quite efficient
in solar flares, it remains unclear whether particle acceleration is
dominated by direct acceleration in current sheets, stochastic
acceleration by turbulence, or shock acceleration (e.g., Miller 1998).

Since global MHD simulations of RIAFs find highly time-dependent
dissipation of magnetic energy (e.g., Hawley \& Balbus 2002;
Machida \& Matsumoto 2003; 
Igumenshchev et al. 2003), it is natural to suppose that solar
flare-like acceleration events happen in the accretion flow close to
the BH.  We therefore focus on this possibility, but discuss a few
more speculative ideas as well.  We assume that there is
(occasionally) enhanced particle acceleration in the inner region of
the RIAF, at $\lsim 10R_s$. In this scenario, the timescale of the
flare will be set by the accretion timescale at $\sim 10R_s$, or the
Alfven crossing time of large scale magnetic loops in this region,
both of which are of order an hour.  Some fraction of the magnetic
energy in the flare will be used to heat the thermal electrons, while some
fraction will accelerate electrons into a power-law distribution of
the form 

\begin{mathletters}
\be
N(\gamma)d\gamma=N_0\gamma^{-p}d\gamma, \hspace{1cm}
 \gamma_{\rm min} \le
\gamma \le \gamma_c,
\ee
\be
N(\gamma)d\gamma=N_0\gamma_c\gamma^{-p-1}d\gamma, \hspace{1cm}
 \gamma_{\rm c} \le
\gamma \le \gamma_{\rm max}, p\ge 1 \ee
\end{mathletters}


\subsection{Synchrotron Emission by Accelerated Electrons}

In our quiescent models of Sgr A*, the magnetic field at $\lsim 10R_s$
is $B \sim 20 $G.  With this magnetic field strength, electrons with
Lorentz factors $\gamma \sim 10^5$ emit synchrotron radiation in {\it
Chandra's} band.  If there is a sufficient number of such electrons,
i.e., if $p$ is small enough, then these electrons will produce a hard X-ray
flare.  We note in this connection that some calculations of
acceleration in current sheets give quite small values of $p$; e.g., Larrabee et
al. (2003) find $p = 1$ while the numerical simulations of Nodes et
al. (2003) give $p = 1.1-1.5$.

The synchrotron cooling time for electrons emitting in {\it Chandra's}
band ($\nu \approx 10^{18}$ Hz) is $t_{cool} \approx 20 \
B_{20}^{-3/2}$ s, much less than the duration of the flare.  Cooling
is thus quite important (Fig. 2) and there is a cooling break in the
emitted spectrum below the X-rays. 
Note that from \S2.2, the hardest
power-law that can be produced by synchrotron emission above the
cooling break is one with a photon index of $\Gamma = 1.5$.

The thick-dashed line in Fig. 6 shows an example of fitting the X-ray
flare using synchrotron emission.  Electrons in a $\approx 10 R_S$
region close to the BH are assumed to be accelerated into a power-law
with $\gamma_{\rm max} \approx 10^6$, $\eta=5.5\%$,
 and $p=1$.  The cooling break is evident at $\nu
\approx 10^{13}$ Hz.  In {\it Chandra's} band the photon index is
$\Gamma = 1.5$, as explained above.  This spectrum is consistent with
the largest flare observed by {\it Chandra}, and is also consistent
with the average spectrum of all the flares observed thus far with
{\it Chandra}: $\Gamma = 1.3^{+0.5}_{-0.4}$ (Baganoff 2003b).

It is important to note that the fraction of the energy in power-law
electrons in the flare model, $\eta = 5.5 \%$, is not that different
from that in our quiescent model, $\eta = 1.5 \%$ (Fig. 1).
Nonetheless, the synchrotron contribution in the X-rays is much larger
because the electron spectral index $p$ is assumed to be smaller ($p =
1$ vs. $p = 3.5$).  Finally, we note from Fig. 6 that during the
flare, only the X-ray flux changes significantly; the radio, submm,
and IR remain essentially constant. This is consistent with the
current observations and is due to the fact that the total number of
lower energy electrons ($\gamma \sim 1-10$) that emit in the radio-IR
is essentially unchanged during the flare.  Of course, the flare may
heat the thermal electrons at the same time that it produces the hard
power-law distribution.  In this case, there could be an increase in
the submm bump flux coincident with the X-ray flare.  In principle,
coordinated observations could help determine the fraction of the
flare energy going into thermal electrons versus power-law electrons.

It is interesting to consider whether the constraint highlighted
above, namely $\Gamma > 1.5$ because of the cooling break, can be
circumvented.  Within the synchrotron model, the only way is by having
a hard electron energy distribution and not having the electrons cool.
This means that the magnetic field should be sufficiently weak that
the cooling time is $\gsim 1$ hour, the flare duration.  This in turn
requires $B \lsim 1$ G in the emitting region.


We have considered various ways of reducing the magnetic strength in
our models.  The most natural way is to decrease the accretion rate.
At fixed $\beta$, $B \propto \dot M^{1/2}$ so a factor of $\gsim 100$
decrease in $\dot M$ is required.  As explained in \S3, however, we
have found it difficult to explain the quiescent spectrum of Sgr A*
with such a low $\dot M$.  Another possibility is that the ``typical''
field strength close to the BH is indeed $\sim 20$ G as estimated
above, but that the emission we see is dominated by a region of much
lower magnetic field strength.

One region where such a low magnetic field is possible is in current
sheets in the accretion flow.  In magnetic reconnection the strength
of the magnetic field in the region where electrons are accelerated is
much lower than in the ambient medium; in fact, $B\sim 0$ at the
center of the current sheet (e.g., Priest \& Forbes 2000). If a
reconnection event is ``strong'' enough to ensure that there is a
large number of accelerated electrons, and if the newly accelerated
electrons can be trapped in the low-B region for an accretion time,
one could in principle obtain X-ray spectra with $\Gamma<1.5$ from
synchrotron emission.  Due to the uncertainties in reconnection
theory, it is unclear whether the above two conditions can be met.

More plausibly, the emission could be dominated by the corona of the
RIAF or an outflow/jet driven from the surface of the accretion flow.
Since the magnetic field strength will decrease with distance from the
midplane of the flow, the cooling break would be less important in
these regions (though the emission still must occur sufficiently close
to the BH to explain the duration of the flare). This model requires
the acceleration of electrons to be less efficient in the bulk of the
RIAF where the magnetic field strength is higher, otherwise that
region would dominate the emission.

The thin dashed line in Fig. 6 shows a concrete example in which
synchrotron emission from a region with $B \approx 0.3$ G produces an
X-ray flare (in current sheets or outflow/jet).  We assume that
electrons have been accelerated with 
$p = 1.2$ and $\eta = 9 \%$.  For
these parameters, the ratio of the number of power-law electrons
to the thermal electrons is only $\sim 10^{-6}$.  Note that there is
no cooling break in the spectrum; the break at high photon energies
is because we took $\gamma_{max} \sim 10^7$.

\subsection{Inverse Compton Emission by Accelerated Electrons}

Synchrotron emission is a viable explanation for the X-ray flares only
if acceleration of very high $\gamma$ electrons is quite efficient.
Alternatively, lower energy electrons can Compton-scatter synchrotron
photons to produce an X-ray flare.  Since the frequency of an
up-scattered photon is $\sim \gamma^2\nu$, the required
$\gamma \sim \sqrt{10^{18}/10^{12}} \sim 10^3$. From Fig. 2, we see
that cooling is not important for such electrons.  Thus, in principle
the spectrum can be much harder than that produced by the
synchrotron models discussed in the previous section.

The dashed line in Fig. 7a shows an example of an IC model for the
X-ray flares, in which we assume that electrons in a region $\approx
2.5 R_S$ in size are accelerated into a distribution with $p = 0.5$.
The model requires $\eta \approx 120\%$ in order to produce the
luminosity of the flare.  This means that about $40-50\%$ of the
electrons in the volume are accelerated into the power-law
distribution.  The model reproduces the observed X-ray spectrum
reasonably well.  It also predicts a factor of a few variability in
the submm and IR.  This is because the accelerated electrons
responsible for the X-ray flare also produce significant synchrotron
emission at lower frequencies.



The value of $p$ in the above model is rather extreme since it
corresponds to a very hard energy distribution, so one would like to
check whether softer energy distributions are also acceptable.  Fig. 7b
compares the above model with another one with $p=1.1$, $\eta=100\%$,
and $\gamma_{max}=630$. This model requires $90\%$ of the
electrons to be in the power-law distribution, which is rather extreme.
The model is also very close to violating the IR limit on the
spectrum.


\subsection{A Two-Phase Medium?}

The spectra of the observed X-ray flares are interestingly close to
that produced by thermal bremsstrahlung emission, $\Gamma \approx 1$.
The problem with invoking bremsstrahlung to explain the observed
rdiation is that, to produce a luminosity of $L_{35} 10^{35}$ ergs
s$^{-1}$ from a sphere of radius $R$, the gas density must be $n
\approx 10^9 L_{35}^{1/2} T_{e,10}^{1/4} (R/10 R_S)^{-3/2}$ cm$^{-3}$,
where $T_{e,10} = T_e/(10^{10} {\rm K})$.  For comparison, the density
in the inner $10 R_S$ for our quiescent model of Sgr A* is $\approx
10^6-10^7$ cm$^{-3}$ (Fig. 2). This makes it difficult to produce an
X-ray flare using bremsstrahlung emission since it would require a
large change in the gas density which would lead to a much larger
increase in the emission at lower frequencies than is observed 
(see, however, Liu \& Melia 2002 who suggest that variations in $T_e$
and $n$ can be appropriately correlated so as to avoid this
difficulty). Another difficulty with a bremsstrahlung interpretation
is that bremsstrahlung emission is dominated by very large radii in
the RIAF, $\sim R_{acc}$, not small radii where the flare must occur
(see \S3 for our discussion of the quiescent emission).

One interesting way out of these difficulties is to consider the
possibility that there is a two-phase medium in the accretion flow
(see, e.g., Nayakshin \& Sunyaev 2003 for related ideas).  If there is a cooler
denser phase embedded in the hot RIAF, it could satisfy the above
density requirement and give rise to bremsstrahlung emission.
Pressure balance with the surrounding RIAF requires $n_b T_b = n_i
T_i$ where $n_i, T_i$ are the number density and ion temperature in
the RIAF, and $n_b, T_b$ are the density and temperature of the cooler
``blob'' (note that we assume thermal pressure balance; magnetic
pressure could, however, be important; e.g., Kuncic, Celotti, \& Rees
1997)


We require $T_b \gsim 10^8$ K in order for the blob to produce hard
X-ray emission observed by {\it Chandra}.  It then follows that if the
size of the blob is $R \sim 5 R_S$ it can produce bremsstrahlung with
the required luminosity, $L \sim 10^{35}$ ergs s$^{-1}$ (one can
readily confirm that the blob is optically thin to free-free
absorption and electron scattering). Note that neither bremsstrahlung 
nor synchrotron emission from the blob will produce any 
significant flux at frequencies lower than the X-ray, so the spectral fit
at those frequencies will not be affected. The duration of the flare is
presumably set by the time it takes the blob to fall into the
black hole from $\sim 10 R_S$.  The ``blob'' in this
solution is notvery small and actually occupies a decent fraction of the
volume of the accretion flow. In addition, almost all of the mass
accretion occurs via the blob, not the hot RIAF.  This means that the
blob probably must be fed by some other source of gas in the external
medium that is not observed by Chandra, e.g. the cold molecular
material invoked by Nayakshin \& Sunyaev (2003; but note that our
two-phase interpretation of the flare is completely different from theirs).

The primary problem with the bremsstrahlung idea is that it is not clear if a cool
phase can be maintained at the required temperature of $\sim 10^8$ K.
Unlike in the ISM, there is no thermal instability in the background
RIAF.  In luminous AGN, a cool optically thick disk with $T \sim
10^5$ K is believed to coexist with a hot optically thin corona with
$T \sim 10^9$ K.  Analogous optically thick material would have $T
\sim 10^3$ K in Sgr A*, much less than that required.  Clearly, more
work on the possible energetics and confinement of a ``cool'' $\sim
10^8$ K two-phase medium is required before this proposal can be
considered a viable explanation for the {\it Chandra} observations.

\section{Summary and Discussion}

New observations of Sgr A* impose strong constraints on theoretical
models for how gas accretes onto the black hole at the center of our
Galaxy.  In this paper, we have investigated how these observations
can be understood in the context of radiatively inefficient accretion
flows.  The high level of linear polarization detected at 230 GHz
constrains the rotation measure through the accretion flow ($\lsim 2
\times 10^6 \ {\rm rad \ m^{-2}}$) and thus puts an upper limit on the
density of gas near the BH.  This argues for an accretion rate much
less than the Bondi rate of $\sim 10^{-5} \mpy$.  A similar conclusion
has been reached by theoretical studies of the dynamics of RIAFs; in
particular, numerical simulations show that very little of the mass
supplied to the accretion flow at large radii actually reaches the
black hole.  Our baseline model has a net accretion rate onto the BH
of $\dot M \approx 4 \times 10^{-8} \mpy$.  Phrased in terms of $\dot
M \propto r^s$ or $\rho \propto r^{-3/2 + s}$ this corresponds to $s
\approx 0.27$ when we normalize the density and thus the accretion rate
to that inferred from {\it Chandra} observations $\approx 1''$ from
the BH.

We have focused on the possibility that there are both thermal and
nonthermal electrons in the accretion flow; this is natural since the
accreting gas in RIAFs is a hot collisionless magnetized plasma. We
calculate the emission from both types of electrons. Mahadevan (1998) and
\"Ozel et al. (1999) showed that the
low-frequency radio spectrum, which was under-predicted in the
original ADAF models (Narayan et al. 1998), can be explained if a small
fraction $\eta$ of the electron energy resides in the power-law tail.
We confirm that result for the RIAF models considered here, and find that
a fraction $\eta \sim 1.5\%$ is sufficient to explain the data (Fig. 1).
If the maximum Lorentz
factor of the electrons is reasonably large, we find that the power-law index of
the electron distribution must satisfy $p \ga 3.5$; otherwise
synchrotron emission from the power-law electrons violates IR and
X-ray (quiescent state) observations (Fig. 4).  The thermal
electrons in our model are responsible for the submm emission and can
account for the observed level of linear polarization
(\S3 \& Fig. 3). This is consistent with 
the work of Melia, Liu \& Coker (2000).

It is interesting to note that there may be independent observational
evidence for nonthermal electrons in RIAFs. McConnell et al. (2000)
combined data from the COMPTEL experiment on {\em CGRO} with data from
both BATSE and OSSE to produce a broadband $\gamma$-ray spectrum for
the hard state of Cyg X-1 (from 50 keV up to $\sim$ 5 MeV). The data
clearly show a hard tail at high energies; this can be explained as
synchrotron from electrons with a {\em steady state} spectral index of
$p=4.5$.  Since cooling of the electrons is undoubtedly important at
such high energies, the spectral index of the {\em injected} electrons
would be $p=3.5$, similar to that found here.  The hard
state of Cyg X-1 can be naturally explained in terms of an ADAF-like
model (Esin etal. 1998). Therefore, even though the luminosities are
very different, the basic physics of the gas flow and particle
acceleration may be similar in Cyg X-1 and Sgr A*.

Nonthermal electrons may also account for the X-ray flares observed by
both Chandra and XMM-Newton from Sgr A*, as proposed by Markoff
et al. (2001).  Specifically, if some of the electrons close
to the BH ($\lsim 10 R_S$) are occasionally accelerated into a
power-law tail with a very hard power-law index ($p \sim 1$, rather
than $p \sim 3.5$ as is required to explain the radio emission),
synchrotron emission or synchrotron self-Compton emission by these
accelerated electrons can produce X-ray flares similar to those
observed (\S4; Figs. 6-8).  In synchrotron models for the X-ray flare,
the required energy in power-law electrons is $\sim 10\%$ of the
available electron thermal energy in the inner $\sim 10 R_S$.  By
contrast, in IC models it is closer to $100 \%$, i.e., nearly all of
the available energy (and electrons) must go into a power-law tail.
This is a somewhat stringent requirement and so we favor the
synchrotron models.  In addition, the upper limits on the IR emission
from Sgr A* are much more difficult to satisfy in IC models because
such a large fraction of the electrons are accelerated (compare
Figs. 6 \& 7); the IR limits require that the maximum Lorentz factor
in IC flares must be small, $\gamma_{\rm max}\la 10^3$, for which
there is no natural explanation.



Our synchrotron flare models in Fig. 6 produce no change in the
radio-IR flux during the flare; this need not, however, be a robust
feature of ``synchrotron'' flares.  If some of the ambient thermal
electrons are heated during the flare (which we have not accounted for
in Fig. 6), they will produce additional submm-IR flux, leading to
correlated radio-X-ray variability.  This could produce variability
similar to that expected in SSC models, making it difficult to pin
down the emission mechanism using variability (though the continued
absence of any lower frequency counterparts to the X-ray flares would,
we believe, favor synchrotron models).
 
In addition to the solar-like flare models, we have also briefly
considered a very different, more speculative, explanation for the
observed X-ray flares, namely the flares could be due to bremsstrahlung emission
from cooler, denser gas embedded in the hot RIAF (\S4.3). Such a model would
nicely explain the very hard X-ray spectra observed (photon index
$\Gamma \sim 1$).  Although in principle viable, this possibility
hinges on an unusual two-phase medium: the ``cool'' gas must still be
quite hot, with $T \sim 10^8$ K, to produce hard X-ray emission (the
background RIAF has $T_p \sim 10^{11}-10^{12}$ K close to the BH).
More work is needed to determine whether this is physical.

One feature of our calculations should be pointed out: the
model shown in Fig. 1 has $\dot M \approx 4 \times 10^{-8} \mpy$ close to the black
hole.  If the outer radius of the RIAF is at $\approx R_{acc} \approx
10^5 \ R_S$, and if we set the density at this radius to the value
measured by {\it Chandra}, then 
it implies $s \approx 0.27$, where $\dot M \propto r^s$.
This value of $s$ is smaller than the values  $ \approx 0.5-1$ found in
numerical simulations of RIAFs (e.g., Hawley \& Balbus 2002;
Igumenshchev et al. 2003).  We have found it difficult to account for
the submm emission of Sgr A* with a larger $s$ and thus a lower $\dot
M$ (Fig. 5).  The reason is that for much smaller $\dot M$, the
electron temperature has to be larger to produce a luminosity
comparable to that observed; the hot electrons then occupy such a
large volume that they significantly overpredict the
$10^{10.5}-10^{12}$ Hz emission (Fig. 5).

There are several possible resolutions of this ``problem'': (1) the
outer radius of the rapidly rotating part of the RIAF may be
$\ll R_{acc}$ (because of low angular momentum at large radii).
For a given gas density close to the BH (from the RM constraint) and a
given gas density at $\approx 1''$ (from {\it Chandra} observations),
a larger value of $s$ would then be required, more consistent with the
simulation results.  (2) The electron thermodynamics could be more
complicated then we consider.  For example, the electron heating
($\delta$) could be stronger in the inner part of the accretion flow
close to the BH ($\lsim 10 R_S$) than it is at larger radii.  In this
case, much lower $\dot M$ is consistent with the data because the
volume occupied by the hottest electrons is relatively small and so
the submm emission is less prominent.

For simplicity, we have taken all of the parameters such as $\alpha,
\beta, \delta, \eta,$ $p$ to be independent of radius.  There is then
not much freedom in the choice of parameter values in our model, with
two exceptions: 1) The value of $p$ (the spectral index of injected
power-law electrons) in the quiescent model is not strongly
constrained; it just has to be $> 3.5$.  2) Similarly, the
maximum Lorentz factor $\gamma_{\rm max}$ of the power-law electrons
in the synchrotron model of flares only has a lower limit ($\sim
10^6$), but it could be quite a bit larger; in the latter case, the
predicted flare spectrum would extend to higher frequencies than shown
in Fig. 6.

Finally, we note that instead of the accretion flow alone
producing all the radiation in Sgr A*, a jet may be responsible
for some of the observed emission (e.g., Falcke et
al. 1993; Falcke \& Markoff 2000).  In particular, Yuan, Markoff, \&
Falcke (2001; 2003) have proposed a coupled jet-RIAF model.
In their model, the underlying accretion flow
is described by a RIAF. Close to the BH, a fraction of the accreting
material is ejected to form a jet (outflows are seen in MHD
simulations of RIAFs, though they are not yet particularly well
collimated).  The physical quantities in the jet are self-consistently
matched onto the underlying RIAF (see Yuan et al. 2001), and the spectral
fit to Sgr A* is satisfactory.  As in other
AGN, the self-absorbed synchrotron emission from the outer part of the
jet fits the low-frequency radio spectrum of Sgr A* (that we have
ascribed to non-thermal electrons in the RIAF in our model); the jet
can also account for the high level of circular to linear polarization
observed at low frequencies (Beckert \& Falcke 2002).  

In jet-RIAF models, the synchrotron emission from the base of the jet
close to the BH dominates over the underlying RIAF and accounts for
the submm-bump (because of effective electron heating in the shock
front at the base of the jet).  In addition, Markoff et al. (2001)
proposed that the X-ray flares observed by {\it Chandra} could be
understood as inverse Compton or synchrotron emission from electrons
heated or accelerated in this region.  Our synchrotron and SSC models are 
quite similar to these ideas, although the geometry is different.


While the jet-RIAF model is successful in explaining observations
of Sgr A*, we have shown in the present paper that a pure RIAF model
is also equally successful.
An important question is to determine which dynamical component is
responsible for which part of the observed emission. One possibility
is to assess whether the coherent magnetic field needed to produce
$\approx 10 \%$ linear polarization in the submm is consistent with
that expected in the accretion flow, or whether it requires an
additional component that could be attributed to the acceleration
region of a jet.  Another promising possibility is simultaneous
multiwavelength observations in the radio-submm and X-ray.  In a jet
one would expect submm variability to lead the radio (if, e.g.,
variability is produced by shocks propagating down the jet), while if
the accretion flow dominates the emission, one would expect little
correlation (if variability is due to local turbulence in the flow) or
that the radio would lead the submm (if variability is due to changes
in the accretion rate).

\acknowledgements F.Y. and R.N. were supported in part by NASA grant
NAG5-10780, NSF grant AST 9820686 and AST 0307433. E.Q. was supported in part by
NASA grant NAG5-12043, NSF grant AST-0206006, and an Alfred P. Sloan
Foundation Fellowship. We thank Fred Baganoff, Jun Lin, Sera Markoff,
and the anonymous referee for useful comments.

\clearpage

\begin{figure}
\plotone{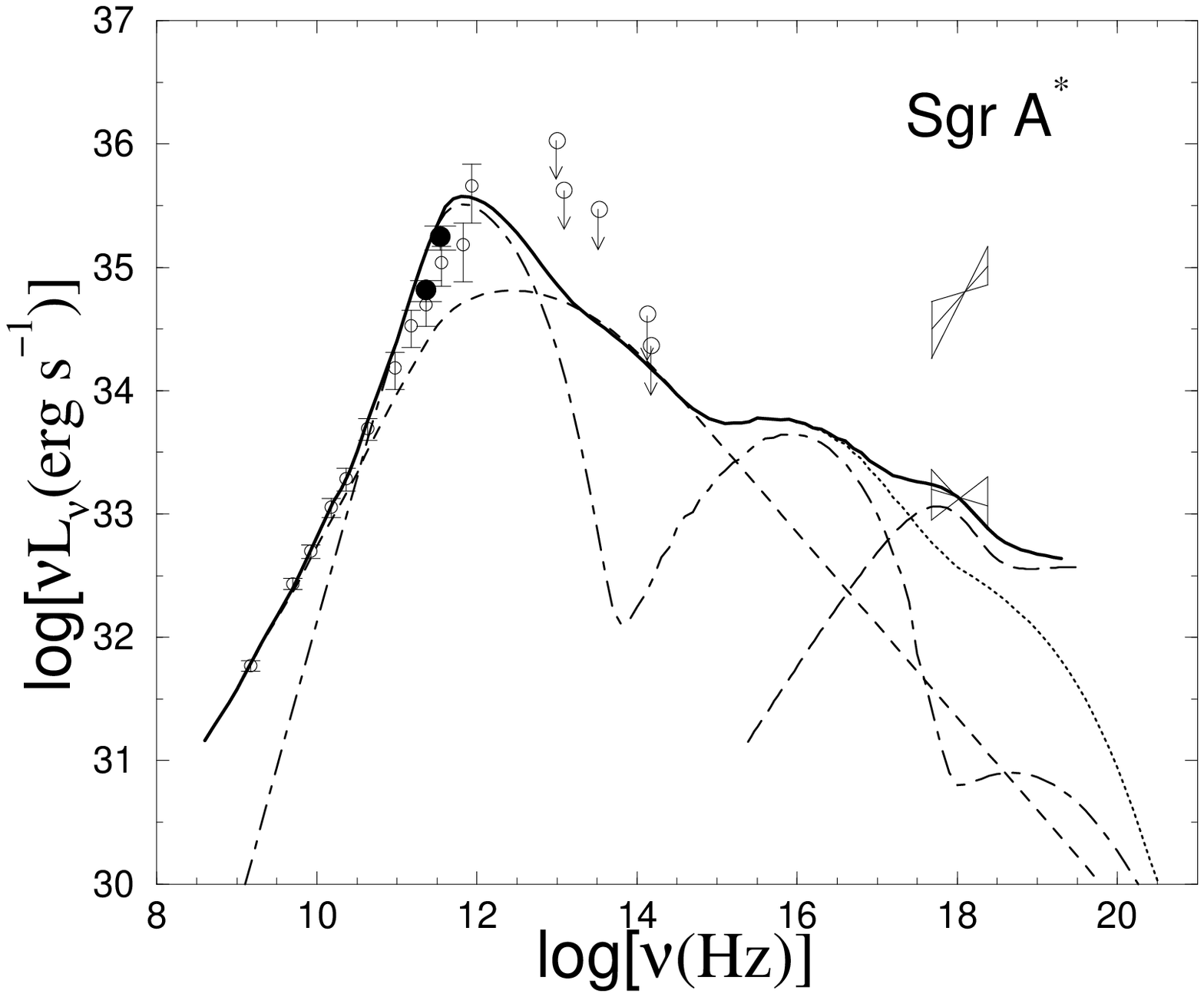}
\caption{A model for the quiescent emission from Sgr A* with $s = 0.27$,
$\delta = 0.55$.  The accretion rate close to the BH is $\dot M
\approx 4 \times 10^{-8} \mpy$.  The dot-dashed line is the
synchrotron and IC emission by thermal electrons. The dashed line is
the synchrotron emission by non-thermal electrons; the non-thermal
electrons have $\approx 1.5 \%$ of the thermal energy with $p=3.5$.
The dotted line is the total synchrotron and IC emission while the
solid line includes the bremsstrahlung emission from the outer parts
of the RIAF (shown by the long-dashed line);  the latter component
explains the extended quiescent X-ray source.  The radio data are from
Falcke et al. 1998 (open circles) and Zhao et al. 2003 (SMA; filled
circles), the IR data are from Serabyn et al. (1997) and Hornstein et
al. (2002). The two ``bowties'' in the X-ray are the quiescent (lower) and
flaring (higher) data from Baganoff et al. (2001; 2003).}
\end{figure}

\begin{figure}
\plotone{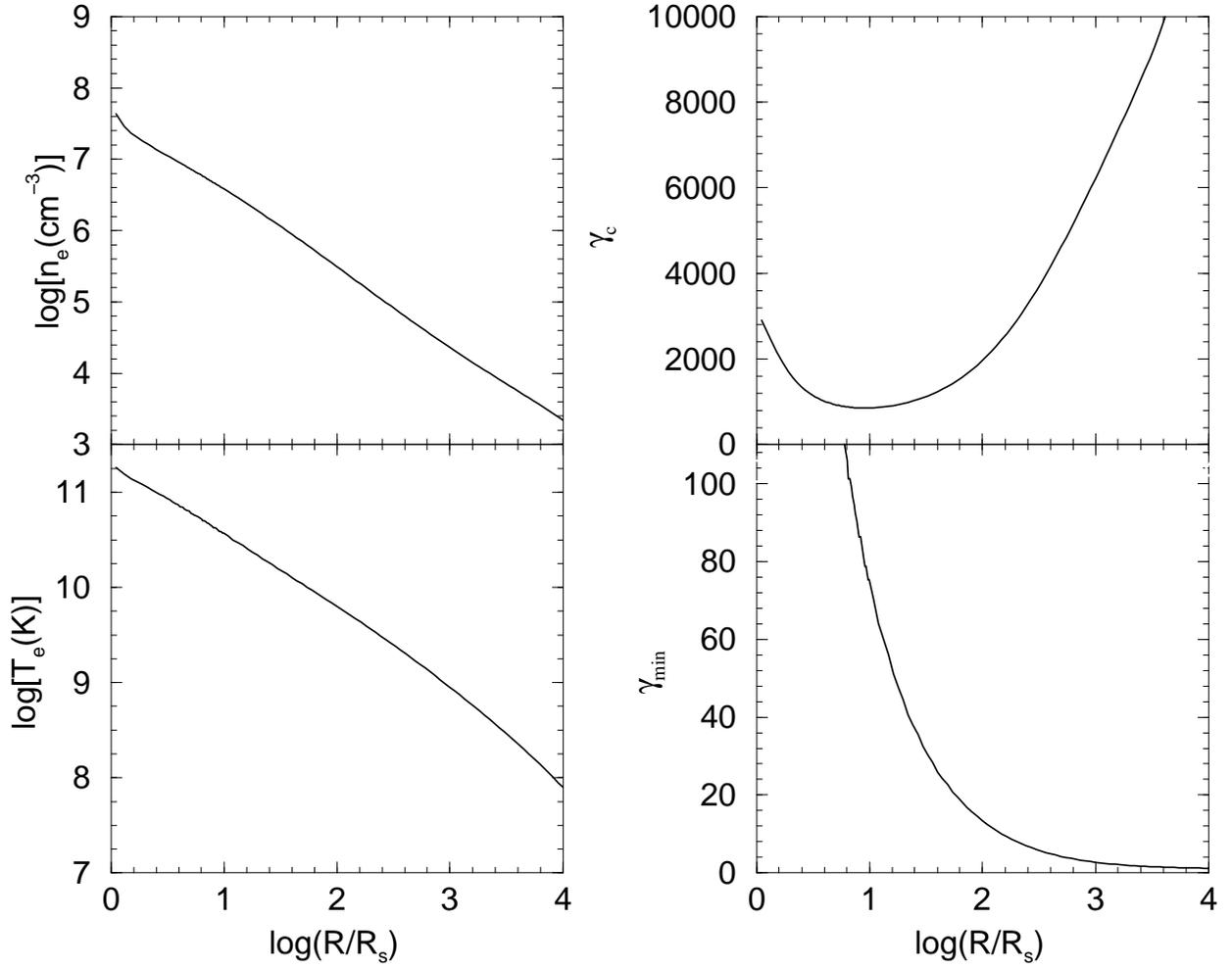}
\caption{The radial profiles of electron density $n_e$, electron 
temperature $T_e$, cooling break Lorentz factor $\gamma_c$, and minimum
Lorentz factor $\gamma_{\rm min}$
for the model shown in Fig. 1.}
\end{figure}

\begin{figure}
\plotone{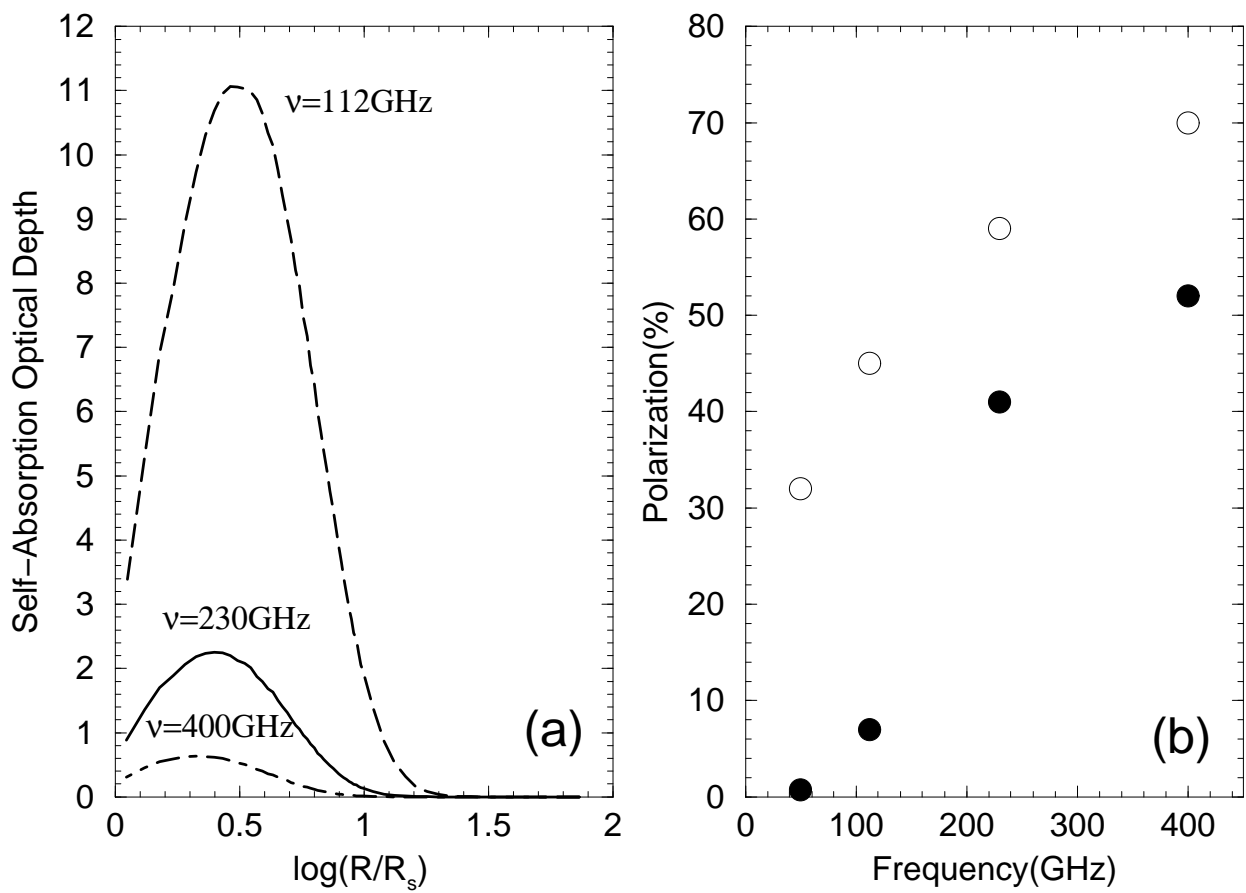}
\caption{(a): Synchrotron self-absorption optical depth of thermal
electrons as a function of radius at three frequencies for the model
shown in Fig. 1. (b): The degree of linear polarization from thermal
electrons at four frequencies with (filled circles) and without (open
circles) Faraday rotation included. These values are upper limits
because they assume a uniform B-field.}
\end{figure}

\begin{figure}
\plotone{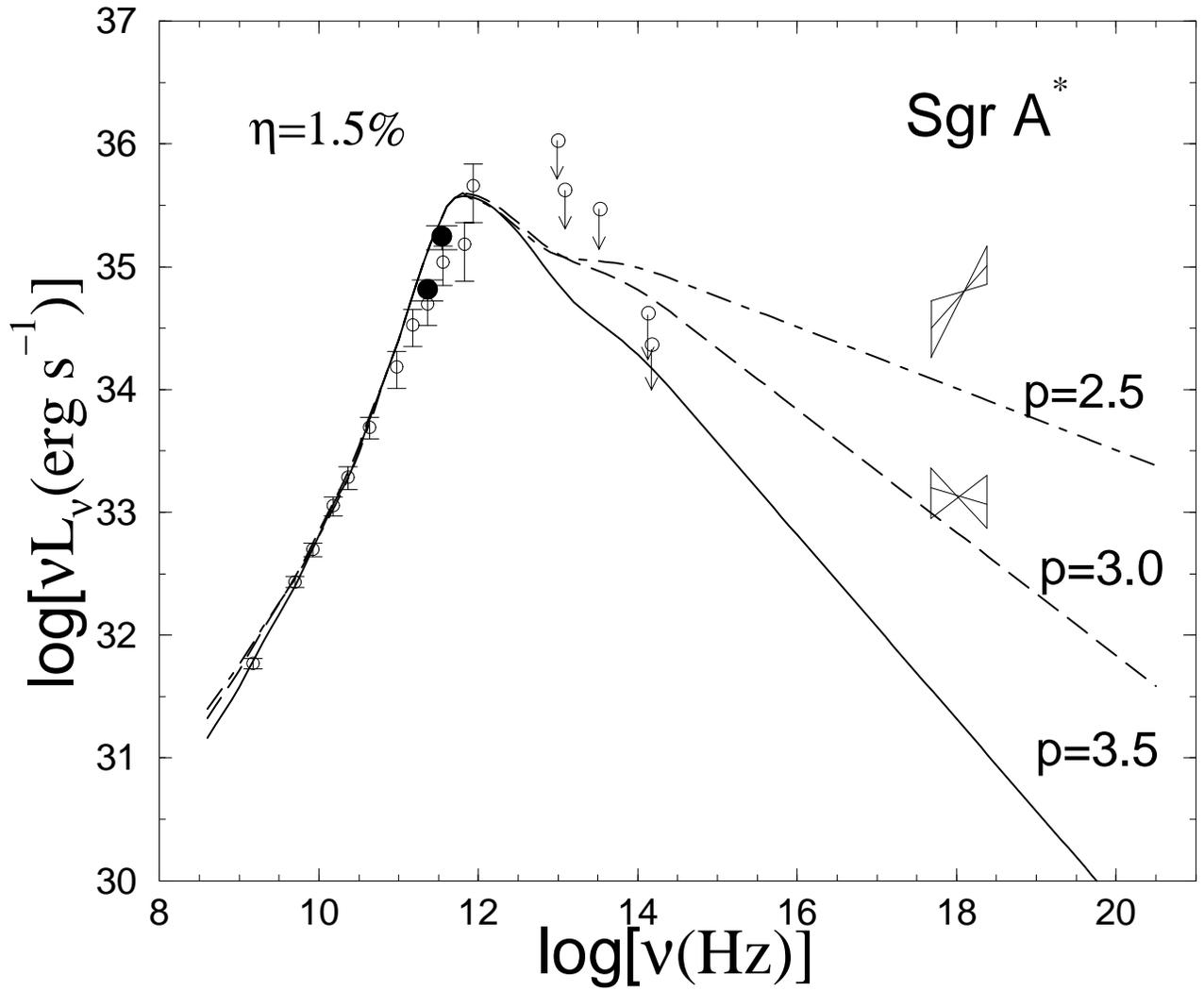}
\caption{The effect of different electron power-law indices $p$ on the
model shown in Fig. 1.  Only the synchrotron emission from power-law and
thermal electrons is shown.}
\end{figure}

\begin{figure}
\plotone{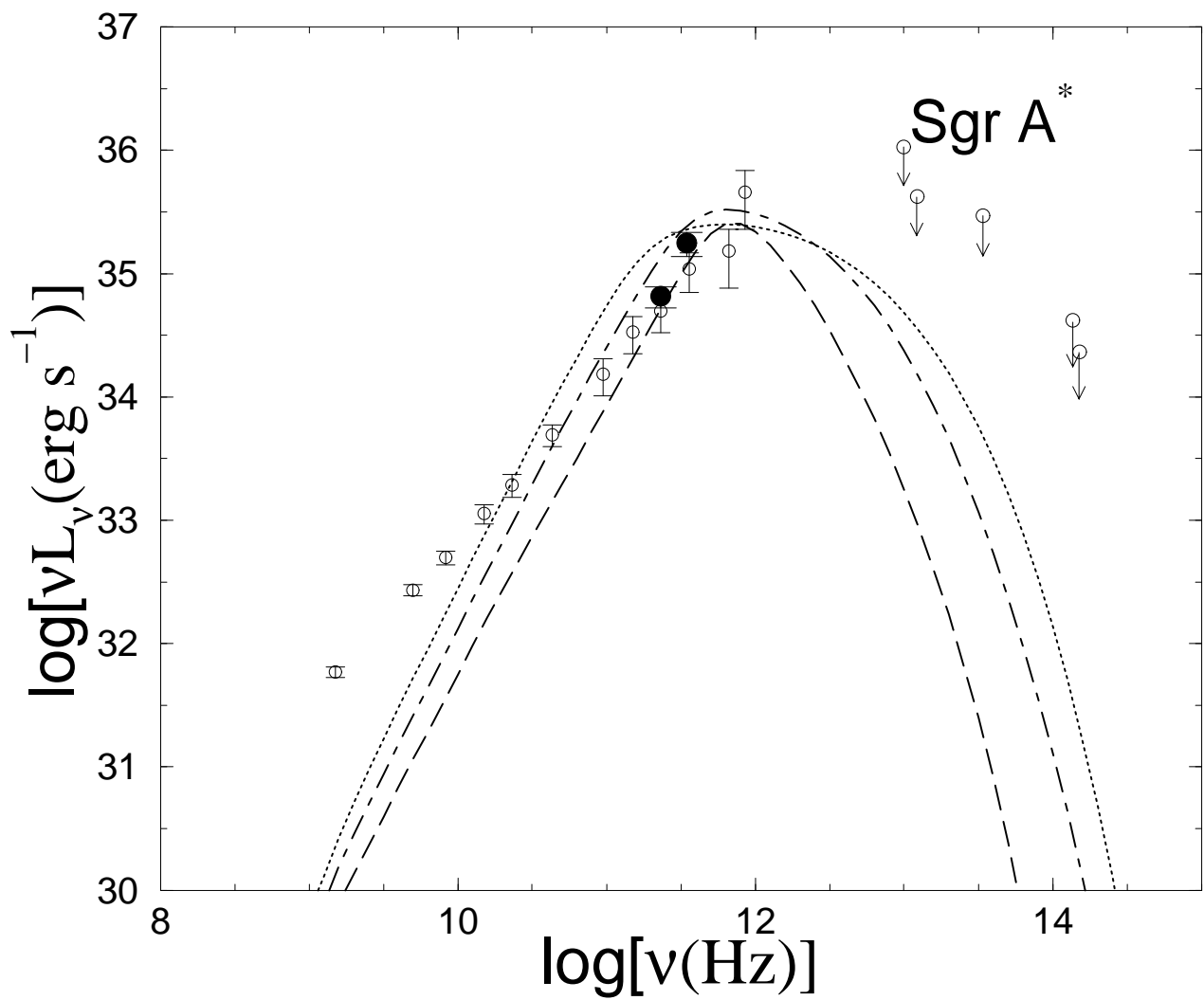}
\caption{The effect of different values of $s$ and $\delta$ on the submm
emission; since the thermal electrons dominate this emission, only
their synchrotron emission is shown.  The models correspond to $s = 0.1$,
$\delta = 0.1$ (dashed line), $s = 0.27$, $\delta = 0.55$
(dot-dashed), $s = 0.4$, $\delta = 1$ (dotted).}
\end{figure}

\begin{figure}
\plotone{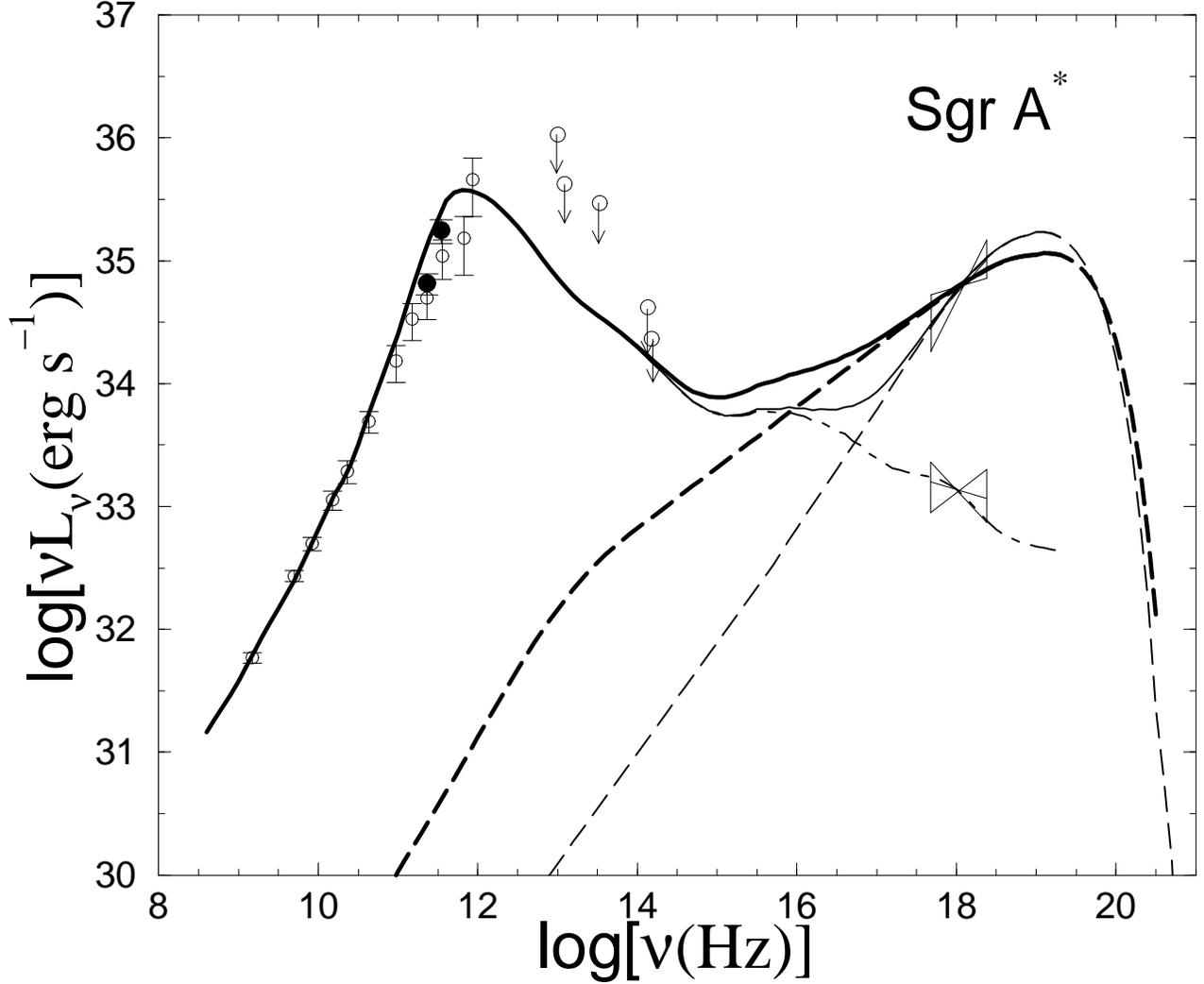}
\caption{Synchrotron models for the X-ray flares from Sgr A*.  The
dot-dashed line is the quiescent emission from Fig. 1.  The thick
dashed line is the synchrotron flare due to power-law electrons
 with $\gamma_{\rm max} \approx 10^6$, $p=1$,
and $\eta = 5.5\%$ (i.e., power-law electrons have $5.5\%$ of the electron
energy); the magnetic field in the RIAF close to the BH is $\sim 20$ G
so there is a cooling break at $\sim 10^{13}$ Hz. The thick solid line
is the total emission during the flare (quiescent emission plus
synchrotron flare).  Finally, the thin lines (dashed and solid) correspond to a
second synchrotron flare model in which $p=1.2$ and the magnetic field
is only 0.5 G so that there is no cooling break.}
\end{figure}

\begin{figure}
\epsscale{0.8}
\plotone{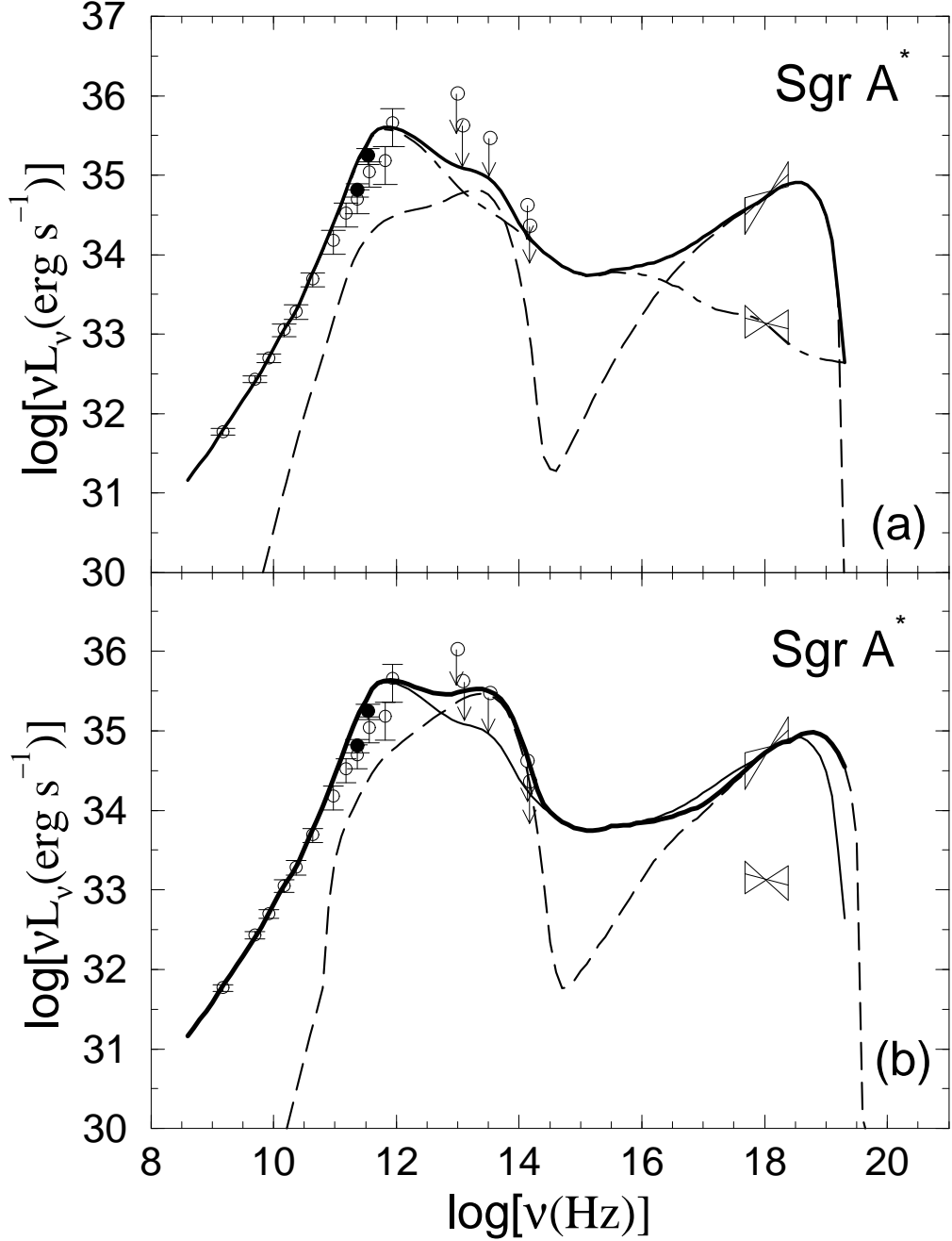}
\caption{Two inverse Compton models for the X-ray flare. (a): The dot-dashed
line is the quiescent emission from Fig. 1.  The dashed line is the
synchrotron and SSC emission in the ``flaring region'' --- power-law
electrons in a $\approx 2.5 R_S$ volume are accelerated with $p=0.5$ and
$\eta=120\%$. The solid line is the total emission during the flare.
(b): The thin solid line is the IC flare model from (a).  The
dashed and thick solid lines are a second IC flare model with similar
parameters except that $p = 1.1$.}
\end{figure}

\end{document}